\documentclass[11pt]{article}
\makeatletter
\renewcommand\section{\@startsection{section}{1}{\z@}%
  {-3.25ex\@plus -1ex \@minus -.2ex}%
  {0.8ex \@plus .2ex}{\reset@font\large\bfseries}}
\@addtoreset{equation}{section}
\renewcommand\subsection{\@startsection{subsection}{2}{\z@}%
  {-3.25ex\@plus -1ex \@minus -.2ex}%
  {0.8ex \@plus .2ex}{\reset@font\normalsize\bfseries}}
\makeatother

\begin{document}
\thispagestyle{empty}
\vspace*{5mm}
\begin{flushright}
\small\sc  December 2000 \\
\small\sc  hep-th/0012018 \\
\end{flushright}
\vspace{20mm}
\begin{center}      
{\Large \bf Singularities of the Seiberg-Witten map     }

\vspace{18mm}
 {\large Andrei G. Bytsko   }

\vspace{5mm}
{\em  Steklov Mathematics Institute\\
Fontanka 27, St.Petersburg, 191011, Russia \\ [2mm]
bytsko@pdmi.ras.ru}
\end{center}
\vspace{1.5cm}

\begin{abstract}
\parbox{12.5cm}{  \vspace*{1mm} \noindent 
We construct an explicit solution of the Seiberg-Witten map for a 
linear gauge field on the non-commutative plane. We observe that 
this solution as well as the solution for a constant curvature diverge 
when the non-commutativity parameter $\theta$ reaches certain event 
horizon in the $\theta$-space. This implies that an ordinary Yang-Mills 
theory can be continuously deformed by the Seiberg-Witten map into 
a non-commutative theory only within one connected component of 
the $\theta$-space.

}\end{abstract}
\vspace*{15mm}

\section{Introduction}

Gauge theories on non-commutative spaces arise as low energy 
effective theories on $D$-brane world volumes in the presence of 
the background Neveu-Schwarz $B$-field in the theory of open 
strings \cite{VS,SW}. The simplest but important example of a
non-commutative space is the space ${\bf R}^{d}_\theta$ which
corresponds to a flat background. The coordinates $x^i$ of 
${\bf R}^{d}_\theta$ obey the Heisenberg commutation relations,
\begin{equation} \label{xx}
  [x^i, x^j] = i \, \theta^{ij} \,,
\end{equation}
where $\theta^{ij}=-\theta^{ji}$ is a constant real-valued
anti-symmetric matrix. Functions on the space ${\bf R}^{d}_\theta$ 
can be identified with ordinary functions on ${\bf R}^{d}$
with the non-commutative product given by the Moyal formula
(here and below summation over repeated indices is implied),
\begin{equation} \label{Mo}
 (u  * v)(x) = \Bigl( e^{ \frac i2 \theta^{ij}  
 \partial_i^x \partial_j^y }\, u(x) v(y) \Bigr)_{y=x} \,.
\end{equation}

As was argued in \cite{SW}, non-commutative Yang-Mills theory on
${\bf R}^{d}_\theta$ is equivalent to ordinary Yang-Mills theory
on ${\bf R}^{d}$. This means that there exists a transformation
relating gauge fields on ${\bf R}^{d}_\theta$ with different values 
of the deformation parameter $\theta$. More precisely, let $\theta$ 
and $\theta+\delta \theta$ be two infinitesimally close values of the 
deformation parameter. Then, there exists the Seiberg-Witten (SW) 
map $A \rightarrow \hat{A}(A)$ and 
$\lambda \rightarrow \hat{\lambda}(\lambda,A)$ of the gauge fields 
and infinitesimal gauge parameters on ${\bf R}^d_\theta$ to those 
on ${\bf R}^d_{\theta+\delta \theta}$ such that
$ \hat{A}(A+\delta_\lambda A)=\hat{A}(A) + 
 \delta_{\hat{\lambda}(\lambda, A)} \hat{A}(A)$.
Explicitly, the SW map is given by \cite{SW}:
\begin{eqnarray} 
 & \delta_\theta A_i =  -\frac 14  \delta\theta^{kl} 
 \{ A_k , \partial_l A_i + F_{li} \}_\theta  \,, & \label{dA} \\
 & \delta_\theta F_{ij} = \frac 14  \delta\theta^{kl} 
 \Bigl( 2\, \{ F_{ik} , F_{jl} \}_\theta - \{ A_k , D_l F_{ij} + 
 \partial_l F_{ij} \}_\theta  \Bigr)  \,, \label{dF}  &
\end{eqnarray}
where  $F_{ij}$ is the curvature of the gauge field defined as follows
\begin{equation} \label{F}
 F_{ij} = \partial_i A_j - \partial_j A_i  - i [ A_i , A_j ]_\theta \,. 
\end{equation}
Here and below $[..]_\theta$ and $\{ ..\}_\theta$ stand, respectively,
for commutator and anti-commutator with respect to the $*$-product.

For slowly varying fields on a $D$-brane, the effective action 
is described by the Dirac-Born-Infeld (DBI) Lagrangian \cite{DBI} 
which depends on $B$, $F$ and the closed string metric $g$.
Using the above equations for $\delta_\theta A_i$ and
$\delta_\theta F_{ij}$, Seiberg and Witten established that
this action is equivalent (up to a total derivative and 
${\cal O}(\partial F)$ terms) to a non-commutative version
of the DBI action. The latter has no explicit $B$-dependence,
but it involves the non-commutative curvature $F(\theta)$ with
%% and all products are of the type (\ref{Mo}) 
\begin{equation} \label{tB}
  \theta (B) = - ( g_\alpha + B)^{-1} B \, (g_\alpha - B)^{-1} \,,
\end{equation}
where $g_\alpha = g/(2\pi \alpha')$,\  $\alpha'$ is the inverse to 
the string tension.

The above mentioned equivalence of two DBI actions is based 
on the idea that one can use the SW map to {\em continuously} 
deform the initial (ordinary) Yang-Mills theory with $\theta=0$ into 
a non-commutative Yang-Mills theory with $\theta=\theta (B)$. 
That is, it is implicitly  assumed that in the $\theta$-space there 
exists a continuous path $\gamma$ connecting the origin with the 
point $\theta=\theta(B)$ such that $A_i$ and $F_{ij}$ converge 
{\em everywhere} on $\gamma$. However,  convergence of 
solutions for the SW map is a rather difficult question which, 
to our knowledge, has not been addressed in the literature. 
In the present paper we will investigate the situation for the 
$U(1)$ constant curvature case. Already in this simplest case 
solutions of the SW map will demonstrate a non-trivial feature --
they cannot be continued beyond certain `event horizon'.

\section{SW map for constant curvature}
\subsection{ Existence of event horizon }
For ${\bf R}^{d}_\theta$, the non-commutativity parameter $\theta$  
has $\frac 12 d(d-1)$ independent entries and thus can be viewed 
as a point in $\frac 12 d(d-1)$-dimensional Euclidean space, which 
we will refer to as the $\theta$-space. Given an initial value of the 
gauge field $A_i$ for $\theta=0$,  we can solve, at least in principle, 
the differential equation  (\ref{dA}) as a series in $\theta$.  As we 
explained above, it is important to determine the region of the
$\theta$-space where this solution converges.

Notice that, in general, solutions of the SW map depend on a path 
in the $\theta$-space along which we carry out the deformation. 
That is, the result of action of two infinitesimal shifts 
$\delta_1 \theta$ and $\delta_2 \theta$ on $A_i$ or $F_{ij}$
depends on their order \cite{AK}. Analogous statement holds for 
a gauge group element $g(x)$ even in  the zero curvature case 
\cite{AB}. However, if  the  curvature is constant (independent of 
the coordinates), its SW map (\ref{dF}) considerably simplifies and 
the corresponding solution does not depend on the deformation path. 
Therefore, we discuss first properties of the solution of eq.~(\ref{dF})
in the $U(1)$ case for constant curvature.

If  the curvature $(F_0)_{ij} $ in the initial (ordinary) theory  is 
independent of the coordinates, then so is  $F_{ij}(\theta)$ 
constructed according to the SW map, which in this case reads 
$\delta F_{ij} =  - \delta\theta^{kl} F_{ik} F_{lj}$. Here we have the 
ordinary product on the right hand side. Therefore, as was noted 
in \cite{SW}, this equation can be rewritten as a matrix differential 
equation (the Lorentz indices are regarded as matrix indices)
\begin{equation} \label{delF}
 \delta F  =  - F \, \delta\theta \, F \,.
\end{equation}
Then the corresponding solution is given in 
the matrix form as follows
\begin{equation} \label{FF}
  F = (1+ F_0 \theta )^{-1} F_0 \,.
\end{equation}
It was remarked in \cite{SW}, that since (\ref{FF}) has a pole at
$\theta=-F_0^{-1}$, there is no non-commutative gauge theory for 
this value of $\theta$ equivalent to the initial commutative theory. 
However, this analysis of equation (\ref{FF}) is not exhaustive. 
Indeed, consider ${\bf R}^{d}_\theta$ and let $F_0$ be 
non-degenerate, i.e., ${\rm rank}\, F_0=d$ (so, $d$ is even). 
Then $F$ diverges for all values of $\theta$ such that 
\begin{equation} \label{detF}
  \det (1+ F_0 \theta ) = 0 \,.
\end{equation}
This equation defines a hypersurface $\Gamma$ in the 
$\theta$-space. To describe $\Gamma$ more explicitly, 
we note  that (\ref{detF}) is solved by the substitution
\begin{equation} \label{tF}
  \theta = - F_0^{-1} + \theta^\prime  \,,
\end{equation}
where $\theta^\prime$ is an arbitrary real-valued $d{\times}d$ 
anti-symmetric matrix such that 
\begin{equation} \label{zd}
 \det \theta^\prime =0 \,. 
\end{equation}
The latter equation describes a cone (as it is invariant 
with respect to the rescaling 
$\theta^\prime \rightarrow {\rm const}\, \theta^\prime$).

An important fact that follows from the above description of
$\Gamma$ is that the complement of  $\Gamma$ in the
$\theta$-space {\em is not} connected. More precisely, this
complement has two connected components. Indeed, the 
complement of (\ref{zd}) consists of all $d{\times}d$ real-valued
anti-symmetric matrices $M$ such that $\det M >0 $.  By an 
orthogonal transformation, $M'=oMo^t$, $o\in O(d)$, any such 
matrix can be brought to the canonical block-diagonal form, 
\begin{equation} \label{cf}
 M' = {\rm diag} ( m_1 \sigma,\, m_2 \sigma ,\, \ldots \,,\, 
 m_{d/2} \sigma)  \,, \quad \sigma = 
 { \textstyle \bigl(    {\,\ 0\atop -1} {1\atop 0}  \bigr)   }
  \,, \quad m_1 \geq m_2 \geq \ldots \geq m_{d/2} >0 \,.
\end{equation}
The group $O(d)$ has two connected components, therefore so 
does the set of all $d{\times}d$ real-valued anti-symmetric matrices 
$M$ with positive determinants. The two connected components
correspond to two possible signs of the Pfaffian ${\rm Pf} M$. 
%% $ = \prod m_i $, if we allow only $SO(d)$ transformations

If $0< {\rm rank}\, F_0=k<d$,  we can apply an orthogonal 
transformation, $F_0'=oF_0 o^t$, such that \hbox{$(F_0')_{ij}=0$} 
for all $i,j>k$. 
Then equations (\ref{FF}) and (\ref{detF}) impose restrictions
only on the upper-left $k{\times}k$-minor of $\theta$.
%%  $$\det\Bigl(\begin{array}{cc} A&B\\C&D\end{array}\Bigr) =
%%  \det A \, \det (D-CA^{-1}B) $$
The remaining $n=\frac 12 (d-k)(d+k-1)$ components of $\theta$
are arbitrary. Thus, in the case of degenerate $F_0$ we have 
$\Gamma= \Gamma_k \times {\bf R}^{n}$, where $\Gamma_k$ 
corresponds to equation (\ref{detF}) for the non-degenerate 
minor of $F_0'$ (and (\ref{detF}) is considered in 
$\frac 12 k(k-1)$-dimensional space).  Consequently, the complements 
of $\Gamma$ and $\Gamma_k$ have the same number of connected 
components, that is two. For  instance,  $\Gamma$  is just 
a single point for $d=2$, hence for $d=3$ we obtain (as $n=2$ here) 
that $\Gamma= {\bf R}^{2}$ is a 2-plane in ${\bf R}^3$.
%% Moreover,  $\Gamma^{d,2}$ is a hyperplane for any $d\geq 2$.
Below we will assume that ${\rm rank}\, F_0=d$ and $d$ is even.

Summarizing, we see that equation (\ref{detF}) defines the 
hypersurface $\Gamma$ whose complement consists of two 
connected components. $\Gamma$ can be viewed as an
{\em event horizon} for the constant curvature version of the
SW map since the corresponding solution (\ref{FF}) diverges 
everywhere on $\Gamma$. As a consequence, the initial 
(ordinary) Yang-Mills theory can be continuously deformed into 
a non-commutative theory only within that connected component, 
call it $\Omega_+$, which contains the origin of the $\theta$-space. 

\subsection{ Estimates on good values of $\theta$   }
The above discussion shows that $\theta$ is `good', i.e., it belongs 
to $\Omega_+$ if the sign of ${\rm Pf}(\theta + F_0^{-1})$ coincides 
with that of ${\rm Pf}(F_0^{-1})$.  Indeed, 
${\rm Pf}(\theta + F_0^{-1})$ vanishes only on $\Gamma$, hence 
its sign is constant within a connected component. Since 
${\rm Pf}(M^{-1})=(-1)^{\frac d2} ({\rm Pf} M)^{-1}$, the condition that 
$\theta$ belongs to $\Omega_+$ is equivalent to the requirement
\begin{equation} \label{sp}
 (-1)^{\frac d2} \, {\rm Pf} F_0 \,  {\rm Pf}(\theta + F_0^{-1})  > 0 \,.
\end{equation}
For instance, in the $d=2$ case we have 
$\theta = \vartheta \, \sigma$, $F_0 = f \, \sigma$ (with $\sigma$ 
as in (\ref{cf})), and (\ref{sp}) is fulfilled if \  $\vartheta f < 1$.  
In the case of $d=4$ any anti-symmetric matrix $M$ can be 
represented by a pair of $3$-vectors, 
$ \vec{e}_M = (M_{12}, M_{13}, M_{14})$,
$ \vec{h}_M = (M_{34}, -M_{24}, M_{23})$. 
In this parameterization ${\rm Pf}\, M=\vec{e}_M \cdot \vec{h}_M$,
so we can rewrite (\ref{sp}) as follows
\begin{equation} \label{eh}
 1+ {\rm Pf} \, \theta \  {\rm Pf} F_0  >  
 \vec{e}_{\theta} \cdot \vec{e}_{F_0}  +
 \vec{h}_{\theta} \cdot \vec{h}_{F_0}  \,.
\end{equation} 
Let us remark that $\Gamma$ for $d=4$ is a cone with the base 
$S^2\times S^2$. Indeed, $\Gamma$ is defined by the equation 
$\vec{e}_M \cdot \vec{h}_M=0$ (with $M=\theta + F_0^{-1}$).
Its intersection with the unit 5-sphere, 
$(\vec{e}_M)^2  + (\vec{h}_M)^2 = 1$, can be parameterized as 
$\vec{e}_M = \vec{a} + \vec{b}$,\  $\vec{h}_M = \vec{a} - \vec{b}$, 
where  $\vec{a}$ and $\vec{b}$ are 3-vectors of length $\frac 12$.

For higher dimensions condition (\ref{sp}) becomes rather involved. 
But it is certainly fulfilled for $\theta$ with sufficiently small 
Euclidean norm, $||\theta||= \sqrt{-{\rm tr}\, \theta^2} $. More 
precisely, if $f_1$ is the maximal eigenvalue for  the canonical form 
(\ref{cf}) of $F_0$, then the distance $r$ between the origin of the 
$\theta$-space and the hypersurface $\Gamma$ is given by (see Appendix)
\begin{equation} \label{min}
 r = \sqrt{2} \, f_1^{-1}  \,.
\end{equation} 
That is any $\theta$ such that $||\theta|| < r$ is guaranteed to belong 
to $\Omega_+$. Note that this estimate does not assume any specific 
form of $\theta$, whereas in the context of the DBI action on a 
$D$-brane we are interested in $\theta(B)$ given by (\ref{tB}).  Here we 
should take into account that the values of $\theta(B)$ belong to some 
{\em compact} domain in the $\theta$-space. To show this, it is 
convenient to introduce new variables,
\begin{equation} \label{gtB}
 \tilde{\theta} = \sqrt{g_\alpha} \,\theta \, \sqrt{g_\alpha} \,, \quad
 \tilde{B} = (\sqrt{g_\alpha})^{-1} \, B \, (\sqrt{g_\alpha})^{-1} \,,
 \quad \tilde{F}_0 = (\sqrt{g_\alpha})^{-1} F_0 (\sqrt{g_\alpha})^{-1} \,,
\end{equation}
where $\sqrt{g_\alpha}$ is defined as a symmetric matrix.
Then (\ref{tB}) acquires the form 
\begin{equation} \label{tB'}
 \tilde{\theta}(\tilde{B}) = - \tilde{B} \,(1- \tilde{B}^2 )^{-1} \,.
\end{equation}
Both sides here are $d{\times}d$ anti-symmetric matrices, so 
this equation defines a map from ${\bf R}^{d(d-1)/2}$ into itself.
This map is continuous because $\det(1-M^2 ) \geq 1$ %%+||M ||^2$ 
for any anti-symmetric $M$ (as follows from (\ref{cf})). Moreover, 
this map is bounded. Indeed, if $b_i $ are the eigenvalues 
corresponding to the canonical form (\ref{cf}) of $\tilde{B}$, 
then the eigenvalues of $\tilde{\theta}$ are given by 
$\vartheta_i= b_i/(1+b_i^2)$. Therefore, the image of the map 
$\tilde{\theta}(\tilde{B})$ (and hence of $\theta(B)$) is a compact 
set in the $\theta$-space. This set belongs to a ball of the radius 
$\frac 12 \sqrt{d}$ (for odd $d$ the radius is $\frac 12 \sqrt{d-1}$ 
since ${\rm rank}\, \tilde{\theta} \leq d-1$). We remark, however, 
that for $d\geq  4$ the boundary of the set is not a sphere since 
the boundary consists of matrices which have canonical forms 
with eigenvalues $\vartheta_1= \frac 12$, 
$0 \leq \vartheta_i \leq \frac 12$,\  $i=2,\ldots, d/2$.

For our purposes, the most important feature of the map 
(\ref{tB'})  is that each eigenvalue of  $\tilde{\theta}(\tilde{B})$ 
has a uniform bound, $\vartheta_i \leq \frac 12$. 
It allows us to state that if
\begin{equation} \label{max}
  \tilde{f}_1 <  2 \,,
\end{equation}
where $\tilde{f}_1$ is the maximal eigenvalue of $\tilde{F}_0$,
then $\theta(B)$ belongs to $\Omega_+$ for any value of the 
$B$-field. To prove this we apply transformation (\ref{gtB}) to 
equation (\ref{tF}). Then, provided that (\ref{max}) holds, the 
canonical form of the r.h.s.~of (\ref{tF}) has at least one 
eigenvalue that is greater than $\frac 12$ (addition of a lower 
rank matrix $\theta^\prime$ can change at most $\frac d2 -1$ 
eigenvalues). Hence the l.h.s.~of (\ref{tF}) cannot be identified 
as $\tilde{\theta}(\tilde{B})$, i.e., the image of the map $\theta(B)$
does not intersect $\Gamma$. This implies that the image,
since it is compact, lies entirely in $\Omega_+$.
 
If equation (\ref{max}) is not fulfilled, then the image of the map 
$\theta(B)$ intersects the event horizon $\Gamma$ (one can 
construct $B$ which solves (\ref{tF}) along the lines of constructing 
$\theta_0$ in the Appendix). In this case a more detailed analysis 
(employing, e.g., (\ref{sp})) is needed to decide whether for given 
$B$ the value of $\theta(B)$ belongs to $\Omega_+$. For instance,
in the $d=4$ case eq.~(\ref{eh}) rewritten in terms of the variables 
(\ref{gtB}) (so that (\ref{tB'}) holds) looks as follows
\begin{equation} \label{ehb}
 1+ \vec{e}_{\tilde B} \cdot ( \vec{e}_{\tilde B} + 
 \vec{e}_{\tilde F_0} ) +  \vec{h}_{\tilde B} \cdot ( \vec{h}_{\tilde B}
 + \vec{h}_{\tilde F_0} ) +  ( \vec{e}_{\tilde B} + 
 \vec{e}_{\tilde F_0} )  \cdot  ( \vec{h}_{\tilde B} + 
 \vec{h}_{\tilde F_0} ) \, {\rm Pf {\tilde B} } > 0 \,.
\end{equation}

\section{SW map for linear gauge fields}
Discussing equivalence of commutative and non-commutative
Yang-Mills theories, it is instructive to study also solutions of 
the SW map (\ref{dA}) for the gauge field $A_i$ and investigate, 
in particular, what singularities they have. However, this task is 
complicated even if the corresponding curvature is constant.
First, the gauge transformations in the non-commutative 
theory are given by \cite{SW}
\begin{equation} \label{gtr}
   \delta_\lambda A_i  = \partial_i \lambda + 
  i [\lambda, A_i ]_\theta \,, \qquad
  \delta_\lambda F_{ij} = i [\lambda, F_{ij} ]_\theta \,,
\end{equation}
where $\lambda$ is the gauge parameter. In the $U(1)$ case a 
constant curvature $F_{ij}$ is gauge-invariant whereas $A_i$ is 
not. Furthermore, a solution of the SW map for the gauge field 
$A_i$ depends on the choice of a deformation path in the 
$\theta$-space even in the zero curvature case \cite{AK,AB}. 

These technical problems are minimized if we consider 
a linear gauge field on ${\bf R}^d_\theta$:
\begin{equation} \label{alin}
    A_i(x) = \alpha_i + a_{ij} \, x^j \,,
\end{equation}    
where $\alpha_i$ and $a_{ij}$ do not depend on the coordinates 
(but, in general, depend on $\theta$). The distinguished feature 
of such a field is that it {\em remains linear} under the SW map. 
This will allow us to solve equation (\ref{dA}) explicitly. 

To commence, we note that, like in the ordinary theory, the 
curvature $F_{ij}$ corresponding to the linear field (\ref{alin}) is 
constant. Therefore, $F(\theta)$ obeys equation (\ref{FF}). On the 
other hand, computing it according to (\ref{F}), we obtain
\begin{equation} \label{Fa}
    F = a^t - a + a \, \theta \, a^t \,,
\end{equation}
where again the Lorentz indices are regarded as matrix indices. Then,
since both $F_{ij}$ and $\partial_i A_j$ are coordinate independent, 
the SW map (\ref{dA}) for the gauge field acquires the form:
\begin{equation}\label{SWa}
 \delta \vec{\alpha} = {\textstyle \frac 12} \, ( a -F) \, 
 \delta\theta \, \vec{\alpha} \,,  \qquad 
 \delta a = {\textstyle \frac 12}\, ( a -F) \, \delta\theta \, a \,.
\end{equation}
Our aim now is to solve this system for given initial data,
$\vec{\alpha}_0 = \vec{\alpha}(\theta=0)$ and $a_0=a(\theta=0)$. 
The latter quantity is not entirely arbitrary but, by equation 
(\ref{Fa}),  is related to the initial curvature, $a_0^t - a_0=F_0$.

The first equation in (\ref{SWa}) does not pose a problem. 
Indeed, we infer {}from the system (\ref{SWa}) that 
$\delta(a^{-1} \vec{\alpha}) =0$. Hence we obtain
$\vec{\alpha} = a \, a_0^{-1} \, \vec{\alpha}_0 $. Further, if $F_0=0$, 
then the second equation in (\ref{SWa}) is also readily solved,
\begin{equation} \label{aF0}
 a = (1 - {\textstyle \frac 12} \, a_0 \theta )^{-1} a_0 \,.
\end{equation}
This formula is consistent with (\ref{Fa}) (which should vanish) 
if we take into account that $F_0=0$ requires that $a_0=a_0^t$. 
Formula (\ref{aF0}) resembles that for the curvature (\ref{FF}). 
However, $a_0$ is now a symmetric matrix, so (\ref{aF0}) has a
different type of $\theta$-dependence. For instance, in the $d=2$ 
case we have $\theta = \vartheta \, \sigma$ (with $\sigma$ as in
(\ref{cf})). Then (\ref{aF0}) becomes
\begin{equation} \label{dF0}
 a(\theta) =   \frac{a_0 + \frac 12 \, \vartheta \, \sigma \det a_0 }%
 {1+\frac 14  \, \vartheta^2 \det a_0 } \,.
\end{equation} 

Let now $F_0\neq 0$. Introduce a new variable $z$ such that
\begin{equation} \label{az}
  a^{-1} = z - F^{-1}  \,.
\end{equation}
Then,  taking into account that $\delta \theta = \delta (F^{-1})$ 
(cf.~equation (\ref{delF})), we rewrite the second equation in 
(\ref{SWa}) as follows
\begin{equation} \label{z}
 2 \, \delta z  \, F = - z \, \delta F  \,.
\end{equation}
It is easy to see from this equation that the result of action 
on $z(\theta)$ of two infinitesimal shifts $\delta_1\theta$ and 
$\delta_2\theta$ depends on their order. More precisely,
\begin{equation} \label{delz}
 [\delta_2, \delta_1] \, z = {\textstyle \frac 14} \,
 [ F \, \delta_1 \theta, F \, \delta_2 \theta ] \,.
%% =\frac 14 [ (\delta_1 F) \, F^{-1}, (\delta_2 F) \, F^{-1}]
\end{equation}
Therefore, in general, to find $z(\theta)$ we have to choose a 
path in the $\theta$-space along which we perform the SW 
transformation. In this context, it is rather an exception (which 
is in agreement with the computations in \cite{AK,AB}) that the 
solution (\ref{aF0}) turns out to be path-independent. 

Nevertheless, we can extract certain information about 
$z(\theta)$ from equation (\ref{Fa}) which now acquires 
the following form
\begin{equation} \label{zz}
   z^t  \, F_0 \, z = - F^{-1} \,.
\end{equation}
In particular, we infer that $(\det z)^2= (\det F_0 \det F)^{-1}$ 
vanishes if $\theta$ belongs to the event horizon $\Gamma$. 
Since $a^{-1} = (1 + z^t  \, F_0 )\, z$, we conclude that $a$ is 
singular everywhere on $\Gamma$. This singularity is 
gauge-invariant as it does not depend on $a_0$.
In addition to this singularity $a$ may have another one
if $\det (1 + z^t  \, F_0)$ vanishes for some values of $\theta$.
This singularity depends on $a_0$ and is not gauge-invariant.

To illustrate this discussion on the behaviour of $z(\theta)$, let us 
present a particular type of solutions to (\ref{z}). Namely, consider
only such paths in the $\theta$-space that $[\delta \theta,F_0]=0$.
Assume for simplicity that all eigenvalues of $F_0$ are different.
Then our requirement ensures that $\delta \theta$ and $F_0$ are 
brought to the canonical form by the same orthogonal 
transformation. This in turn implies that 
$[\delta \theta,\theta]=[F_0,\theta]=0$.  The later equation 
defines a linear subspace of the $\theta$-space. On this 
subspace (\ref{delz}) vanishes and (\ref{z}) is solved by  
$z(\theta)= z_0 \sqrt{1+ F_0 \theta}$, where the square 
root is defined as a real-valued symmetric matrix. The 
integration constant $z_0$ is determined from (\ref{az}).
Thus, we obtain a solution of (\ref{SWa}):
\begin{equation} \label{aF1}
 \begin{array}{rl}
   a^{-1} =  & (a_0^{-1} + F_0^{-1}) \, \sqrt{1+ F_0 \theta }  
   - F_0^{-1} - \theta \\ [2mm]
   = & (a_0^{-1} + F_0^{-1}  - F_0^{-1}  \sqrt{1+ F_0 \theta } 
   ) \, \sqrt{1+ F_0 \theta } \,.
 \end{array}
\end{equation}
It is easy to verify (again taking into account that $a_0^t - a_0=F_0$)
that  this solution is consistent with (\ref{Fa}).  Notice also that 
(\ref{aF1}) turns into (\ref{aF0}) when $F_0$ goes to zero.

A linear gauge field corresponding to the solution (\ref{aF1}) 
becomes singular everywhere on the hypersurface $\Gamma$ 
and besides it diverges if 
$\det ( a_0^{-1} + F_0^{-1} (1- \sqrt{1+ F_0 \theta } ) ) =0$. Actually, 
since (\ref{aF1}) contains the square root of $(1+ F_0 \theta)$, this 
solution makes sense only in the connected component $\Omega_+$. 
Indeed, when $\theta$ passes through the event horizon $\Gamma$, 
some of the eigenvalues of $(1+ F_0 \theta)$ become negative, and 
$\sqrt{1+ F_0 \theta }$ cannot be defined appropriately. It is plausible
that this picture holds for all solutions of (\ref{SWa}) since, as 
seen from (\ref{z}), they unavoidably involve some kind of a square 
root of $F_0F^{-1}$. This demonstrates again that the initial 
(ordinary) Yang-Mills theory can be continuously deformed into 
a non-commutative theory only within the connected component
$\Omega_+$ of the $\theta$-space.

In the $d=2$ case the $\theta$-space is one-dimensional and
$\theta = \vartheta \, \sigma$ commutes with $F_0 = f \, \sigma$
(here $\sigma$ is as in (\ref{cf})). Therefore, in this case (\ref{aF1})
gives the general solution. In the explicit form, it looks as follows
\begin{equation} \label{aa}
  a(\theta) = \frac{f^2} {\gamma_\theta \beta_\theta } \ a_0 + 
  \frac{f (1-\gamma_\theta) \det a_0} {\gamma_\theta \beta_\theta } 
  \  \sigma \,,
\end{equation}
where $\gamma_\theta = \sqrt{1- \vartheta f} $ and
$\beta_\theta = f^2  \gamma_\theta + (1- \gamma_\theta)^2 \det a_0$.
For example, if the gauge field in the ordinary Yang-Mills theory 
has the form $A_1=- \frac 12 x^2$, $A_2= \frac 12 x^1$, then it will 
evolve under the SW map into
\begin{equation} \label{ex}
 A_1(\theta) = \frac{- x^2}{(1+\sqrt{1-\vartheta })
 \sqrt{1-\vartheta}}  \,, \quad
 A_2(\theta) = \frac{ x^1}{(1+\sqrt{1-\vartheta })
 \sqrt{1-\vartheta}}   \,. 
\end{equation}

Finally, it is interesting to remark that if $\det a_0 =0$, then 
(\ref{aa}) simplifies to $a(\theta) = (1- f \vartheta)^{-1} a_0$, i.e., to
the same dependence on $\theta$ which the curvature $F(\theta)$
has (cf.~(\ref{FF})). However, it is not clear what is an analogue of 
this observation for higher dimensions $d$.

\section{Conclusion}
We have considered solutions to the SW map in the constant
curvature case and observed that they diverge when $\theta$
reaches certain event horizon $\Gamma$ which divides the
$\theta$-space into two connected components. This implies that 
an ordinary Yang-Mills theory can be continuously deformed into a 
non-commutative theory only within the connected component
$\Omega_+$ which contains the origin of the $\theta$-space. We have 
found some sufficient conditions for $\theta$ to belong to $\Omega_+$
which involve only the maximal eigenvalue of the initial curvature 
$F_0$. These results can be of interest in the context of the 
deformation quantization approach to gauge theories on 
non-commutative manifolds. In particular, one can conjecture that 
there exist conditions on $\theta$ ensuring convergence of the SW map 
in the general case in terms of the maximal eigenvalue of the 
corresponding curvature $F_0(x)$.

{}From the string theory viewpoint, our results indicate that 
equivalence of the ordinary and non-commutative DBI actions holds 
possibly not for all values of the $B$-field. Although we studied here 
only the constant curvature case, we expect that the event horizon 
for the SW map exists also in a non-constant case (at least, for slow 
varying fields).  A supporting evidence for this is provided by the 
formula relating $F_0(x)$ and $F(x)$ recently suggested in~\cite{Liu}.  
%% it contains $\sqrt{\det (1-\theta F)} (1- F \theta)^{-1} F$ 

In the present paper we have constructed an explicit solution of the 
SW map for a linear gauge field (apart from formula (\ref{FF}) this is, 
to our knowledge, the first exact solution to the SW map). It possibly 
can be useful for further development of the perturbative approach to
solving the SW map \cite{Liu,sol}. In particular, our solution for the
gauge field and the solution for the curvature can be written as follows
\begin{equation} \label{sym}
 F^{-1}  - F_0^{-1} = \theta \,, \qquad
 a^{-1} + F^{-1} = (a_0^{-1} + F_0^{-1}) \, \sqrt{F_0 F^{-1}} \,. 
\end{equation}
This form of the solutions exhibits apparent symmetry. Namely, 
(\ref{sym}) remains invariant if we reverse the sign of $\theta$ and, 
at the same time, exchange the ordinary and the non-commutative 
variables (recall that here $F$ and $F_0$ commute because of the 
specific choice of a subspace in the $\theta$-space), 
\begin{equation} \label{rev}
 \theta \leftrightarrow  - \theta \,, \qquad 
 F \leftrightarrow  F_0 \,, \qquad  a \leftrightarrow  a_0 \,.
\end{equation}
It would be very interesting to see whether solutions
of the SW map  in more general cases possess similar 
symmetries.

\vspace*{3mm}
\noindent {\bf Acknowledgements:} \
I thank A. Alekseev for valuable discussions, and G. Perelman 
and V. Roubtsov for useful comments. This work was supported 
in part by the INTAS grant 99-01705 and by the grant VISBY-380 of
the Swedish Institute.

\section*{ Appendix }
Let us find the distance $r$ between the origin of the $\theta$-space 
and the hypersurface $\Gamma$. That is, we want to find the 
minimum of the Euclidean norm, $||\theta||= \sqrt{-{\rm tr}\, \theta^2}$,
for $\theta$ which solve equation (\ref{detF}).  Let an orthogonal 
transformation with some $o \in O(d)$ brings $F_0$ to the 
canonical form (\ref{cf}) with eigenvalues 
$f_1 \geq f_2 \geq \ldots \geq f_{d/2} >0$. 
Then, clearly, $o^t \theta_0 \, o$ solves (\ref{detF}) if we take
$\theta_0= {\rm diag}( f_1^{-1} \sigma \,,\, 0\,,\, \ldots \,, 0)$. 
In fact, $ ||\theta_0|| = \sqrt{2} f_1^{-1} $ is {\em the} minimum which 
we are seeking. To prove this assertion, we notice that, since 
${\rm rank}\, \theta^\prime \leq d-2$ in (\ref{tF}), we can apply to 
this equation such an orthogonal transformation with some 
$o \in O(d)$ that $o\, \theta^\prime o^t$ acquires a form where 
entries outside of the upper-left $(d-2){\times}(d-2)$-minor vanish. 
Then, as we want to minimize $||\theta||$, we should put 
$o \, \theta^\prime o^t$ equal to the upper-left 
$(d-2){\times}(d-2)$-minor of $o F_0^{-1} o^t$. Since $||..||$ is 
invariant with respect to orthogonal transformations, we conclude 
that $||\theta||^2 \geq ||F_0^{-1}||^2 - ||\hat{F}_0^{-1}||^2$, where 
$\hat{F}_0^{-1}$ is any $(d-2){\times}(d-2)$-minor of $F_0^{-1}$. 
Let $\lambda_1 \leq \lambda_2 \leq \ldots \leq \lambda_{\frac d2 -1}$ 
be the eigenvalues (each of the multiplicity two) of $-\hat{F}_0^{-2}$. 
Taking into account that $-{F}_0^{-2}$ is a positive definite matrix, we 
can apply a theorem {}from the linear algebra (see, e.g., % Theorem 14 in
chapter 10 of \cite{Gan}), stating that $\lambda_i$  satisfy the 
relation $f_{i}^{-2}\leq \lambda_i \leq f_{i+1}^{-2}$. Therefore,
$||\hat{F}_0^{-1}||^2 \leq 2 \sum_{i=2}^{d/2} f_i^{-2}
 = ||F_0^{-1}||^2 - 2 f_1^{-2} $ and hence we proved for equation 
(\ref{tF}) that $||\theta||^2 \geq 2 f_1^{-2}$. Thus, $ ||\theta_0|| $ 
found above is the minimal possible value, i.e., it is the distance 
$r$ between the origin of the $\theta$-space and $\Gamma$.

\newcommand{\sbibitem}[1]{ \vspace*{-1.5ex} \bibitem{#1}  }

\end{document}